# Overcomplete Wavelets for Compressed Sensing

Bhabesh Deka

Department of Electronics and Communication Engineering
Tezpur University, Napaam, Tezpur, Assam, India 784028

**Abstract.** Compressed sensing (CS) using overcomplete wavelet dictionaries has been a well-investigated topic in the recent times for image and vision applications. In this paper, different overcomplete wavelet transforms have been studied to estimate the best transform. Performance evaluations are carried out for different overcomplete wavelet transforms from highly undersampled and inaccurate measurements for the recovery of images in frequency as well as physical domains.

## 1 Introduction

Compressed sensing (CS) aims to recover a signal, which in some way difficult to measure, but otherwise naturally gifted with the unique property called the *sparsity* or compressibility over another domain quite different from its physical domain by measuring only a few projections of the signal instead of its samples directly. This is far less than the Shannon-Nyquist sampling rate [1]. Success of CS heavily relies on the existence of the transform or dictionary over which the signal is having the best possible representation in a *sparse* way. Besides such transform or dictionary must also be nearly orthogonal or *incoherent* to the signal projection or measurement matrix, which in some cases are already fixed (like in the case of magnetic resonance imaging (MRI)) or in others, have limited flexibility due to hardware limitations (wireless sensors in body area networks). So, from the signal processing point of view choosing the best transform or design of the overcomplete dictionary is of paramount importance or interest for efficient and exact recovery of the signal from the available measurements. As reported in [1], sparsity gives the representation of a signal as a linear combination of a few large coefficients and coherence is the measure of maximum correlation between measurement and representation bases. Discrete wavelet transform (DWT) is the sparsifying overcomplete wavelet transform that has been in use for signal processing for many years with many successful and breakthrough applications. DWT can give optimal sparse representation for signals, which are piecewise smooth and have singularities, like, jumps and spikes. It is because wavelets are very compact functions and singularities (due to discontinuities in the signal) produce large magnitude wavelet coefficients, which are distinct from others. However, the DWT suffers from some fundamental limitations, like, oscillations near discontinuities, shift-invariance, lack of directional property, and aliasing due to downsampling operation. The double-density DWT (DD-DWT) [2] and the dual-tree complex DWT (DT-CoWT) [3] are redundant or overcomplete transforms (having a redundancy factor of two), nearly shift-invariant, and based on FIR perfect reconstruction filter banks. DD-DWT approximates the continuous wavelet transform by introducing additional wavelet function while the DD-CoWT possesses special properties of complex wavelet functions suitable for vision and image information processing. Both DD-DWT and DD-CoWT outperform the critically sampled DWT. Moreover, DD-CoWT is a complex-valued wavelet transform and most suitable for signal modeling and denoising.

Double-density dual-tree DWT (DD-DT-DWT) [4] is an overcomplete transform and possesses merits of both DD-DWT and DD-CoWT i.e. more wavelet functions to approximate properties of the continuous wavelet transform and complex wavelets functions for an effective image representation. As reported, the DD-DT-DWT overcomplete dictionary will be more suitable for image denoising, enhancement, and segmentation and most importantly in sparse signal representation.

In this paper, our aim is to study the incoherence properties of different overcomplete wavelet transforms and evaluate suitability of the transform for compressed sensing applications. In particular, we study the restricted isometric property (RIP) for these overcomplete wavelet dictionaries with sensing matrices in physical and frequency domains. Next, performances of these overcomplete wavelet transforms are seen for compressed sensing image reconstruction from inaccurate and highly undersampled measurements in both domains.

Rest of the paper is organized as follows: in Section 2 a brief background on overcomplete wavelets and important definitions on CS are given. Section 3 discusses on the methodology adopted and Section 4 gives detailed simulation results. Finally, conclusions are drawn in Section 5.

## 2 Background and Definitions

In this section, a short theoretical background on overcomplete wavelet transform is discussed. DWT is based on two multiresolution expansion functions that is the scaling and wavelet functions. 2-D DWT consists of one 2-D scaling function and three 2-D wavelet functions. Fig. 1 shows the 2D wavelet functions. The HH wavelet subband does not have the ability to isolate orientations as it mixes the $-45^0$ and $+45^0$ orientations. Lack of directionality and other limitations: shift-variance, aliasing, oscillations are the main problems in the DWT. In order to overcome these limitations, many researchers in signal processing later developed *overcomplete wavelet transforms* or *frames*.

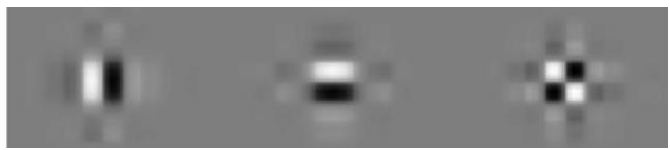

Figure 1: Orientation of LH, HL, HH wavelets respectively

## 2.1 Dual Tree CoWT (DT-CoWT)

Motivated from the fact that Fourier transform does not oscillate between positive and negative values near the discontinuity, is perfectly shift-invariant, does not give aliases, and highly directional, the dual-tree CoWT was conceived in [3]. It is designed by interconnecting two DWTs, one in the real part (upper half) and the other in the imaginary part (lower half). In Fig. 2 (a), low-pass filters, $h_0(n)$ and $g_0(n)$ maintain one-half sample shift such that the two wavelets (in both the halves) form a Hilbert transform pair:

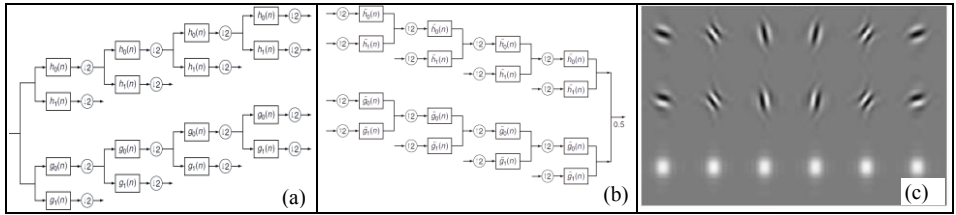

Figure 2: Filter bank structures: (a) forward DT-CoWT , (b) inverse DT-CoWT (source Figs. 7-8 [3]) and (c) Wavelet functions of 2-D DT-CoWT (Source: Fig. 16 [3])

$$\psi_g(t) \approx H\{\psi_h(t)\}, \quad (1)$$

where $\psi_h(t)$ and $\psi_g(t)$ are the two real wavelets. A complex wavelet function is obtained by combining them, which is also approximately analytic, i.e. supported on one-half of the frequency axis only. That is,

$$\psi(t) = \psi_h(t) + j\psi_g(t). \quad (2)$$

In Fig. 2(c), first row indicates wavelet subbands orientations in the real part, the second row shows the same in the imaginary part and third row shows the magnitude. Thus, it is expansive by a factor of 2 for 1-D and by a factor of 4 for 2-D that is $2^d$ for a $d$-dimensional signal.

## 2.2 Double Density DWT (DD-DWT)

DD-DWT is nearly shift-invariant like that of the DT-CoWT and has a redundancy factor of 2. It contains one scaling function and two wavelet functions shifted by one-half from one another [2]:

$$\psi_2(t) \approx \psi_1(t - 0.5) \quad (3)$$

The block diagram of forward and inverse DD-DWT including analysis and synthesis filters is shown in Fig. 3 (a). This transform has more degrees of freedom in comparison to DT-CoWT as now the wavelet functions need not be a Hilbert Transform pair. This further means that they lack directionality property of DT-CoWT. Therefore, to preserve the merits of DT-CoWT, two DD-DWTs are merged to form a new overcomplete wavelet transform that is the double-density dual-tree DWT

(DD-DT-DWT) [4].

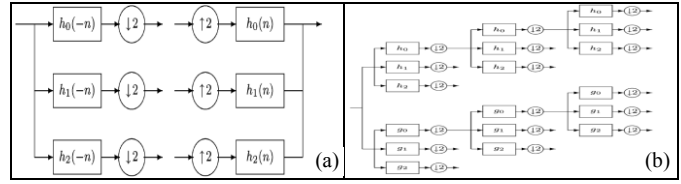

Figure 3: Filter bank structures in (a) forward and inverse DD-DWT (b) forward DD-DT-DWT

## 2.3 Double-Density Dual-Tree DWT (DD-DT-DWT)

It has two scaling functions and four wavelet functions, that is $\psi_{h,i}(t)$ and $\psi_{g,i}(t)$ for $i = 1, 2$. These wavelets also contain one-half shift property such that:

$$\psi_{h,1}(t) \approx \psi_{h,2}(t-0.5), \tag{4}$$

$$\psi_{g,1}(t) \approx \psi_{g,2}(t-0.5) \tag{5}$$

The above wavelet functions form Hilbert transform pairs, i.e.

$$\psi_{g,1}(t) \approx H\{\psi_{h,1}(t)\}, \tag{6}$$

$$\psi_{g,2}(t) \approx H\{\psi_{h,2}(t)\} \tag{7}$$

Fig. 3(b) shows the filter bank structure for analysis. It is expanded by a factor of 4 in 1-D and in 2-D, it consists of 32 oriented wavelets and have the directionality property.

## 2.4 Sparsity and Incoherence

Most of the signals in the nature are sparse when represented over a proper basis set $\psi^T$. Thus, a sparse signal $f$ can be represented as: $f = \psi^T x$, where $x$ is a compressible or dense signal. CS guarantees successful signal recovery, if the sensing basis, $\phi$ is incoherent to the representation basis, $\psi$. The expression for mutual coherence $\mu$ between two orthobasis pair, $\phi$ and $\psi$ is given by [1], i.e.

$$\mu(\phi,\psi) = \sqrt{N} \max_{1 \leq k,j \leq N} |\langle \phi_k, \psi_j \rangle|, \tag{8}$$

where $\phi_k$ and $\psi_j$ are columns of $\phi$ and $\psi$, respectively; $N$ is the length of the signal. Consider the CS measurement matrix $A$ of size $M \times N$ ($M \ll N$),

$$A = R\phi\psi^T, \tag{9}$$

where $R$ is a binary matrix of size $M \times N$ that selects $M$ rows from $\phi$. The restricted isometry property (RIP) states that for a $S$-sparse vector, $f$ the $S$-columns taken from $A$ need to be nearly orthogonal, mathematically,

$$(1-\delta_S)\|f\|_2^2 \leq \|Af\|_2^2 \leq (1+\delta_S)\|f\|_2^2 \tag{10}$$

## 3 Proposed Methodology

In order to carry out CS reconstructions with overcomplete wavelet transforms successfully, it is desired that transforms should have maximal incoherence with the sensing matrix. Therefore, in the proposed work we first estimate the mutual coherence of different overcomplete wavelet transforms with a relatively simple procedure before applying the CS reconstruction algorithm.

### 3.1 Estimation of mutual coherence

Approximation of mutual coherence of the overcomplete wavelet transform may be done by Monte-Carlo (MC) simulations. Instead of finding the full set of inner products as in Eqn. (8), the mutual coherence can be approximated for a subset of columns of the CS matrix $A = WRF\psi^T$ defined by:

$$v_i = \max |A^T A|, \qquad (11)$$

where $\psi$ is one of the overcomplete wavelet transforms discussed above, $F$ the Fourier transform matrix; representing $\phi$ in the frequency domain in Eqn. (9), $R$ is the mask operator that applies 50% random undersamplimg to the frequency domain data of the test image, and $W$ is the diagonal matrix such that norm of $A$ is 1. The basis function for the transform $\psi$ is estimated by taking the inverse wavelet transform on a matrix having only single non-zero value at a random location in the HH-subband. Examples from different transforms for the HH-band are displayed in Fig. 4. Next, we repeat the procedure for other random positions of the non-zero value in the same subband and calculate $v_i$. Finally, the estimate $\tilde{\mu}$ between $\psi$ and $\phi$ is calculated as: $\tilde{\mu} = \max\{v_i\}_{i=1,2,\cdots,L}$, where $L$ is a big number. We adopt a strategy like the one applied in [5, 6] for the estimation of mutual coherence.

### 3.2 CS reconstruction from partial measurements

If $x$ is sparse, CS reconstruction is able to recover the sparse vector $x$, given its measurements $y$ by solving the following constrained optimization problem, i.e.

$$\min_x \|x\|_{l_1} \text{ subject to } y = R\phi x + \eta, \qquad (12)$$

where $R^{m \times n}$ is the binary matrix having exactly one non-zero value in every row at random locations, $\phi^{n \times n}$ is the sensing basis and $\eta$ is the additive white Gaussian noise vector. Since elements $x_i$ are independent, so we can solve an unconstrained minimization problem for each $x_i$, i.e.

$$\arg\min \frac{1}{2}|Ax_i - y_i|^2 + \lambda|x_i|, \qquad (13)$$

where $\lambda$ is a small positive regularization parameter or threshold and $A = R\phi$. The solution $\hat{x}$ is obtained by:

$$\hat{x} = \begin{cases} y + \lambda & \text{if } y < -\lambda \\ 0 & \text{if } |y| < \lambda \\ y - \lambda & \text{if } y > \lambda \end{cases} \quad (14)$$

Eqn. (14) is also called the soft thresholding or shrinkage function $S(\hat{x}_i, \lambda)$. However, for all practical signals, the above assumption on $x$ is overly simplistic. Therefore, we consider that although $x$ is not sparse in practice, yet it is compressible or has a sparse representation over a different basis set $\psi^T$ i.e. $\psi^T x$ is sparse. This leads to the modification of the Eqn. (12) as follows:

$$\min_x \|\psi^T x\|_{l_1} \text{ subject to } y = Ax + \eta \quad (15)$$

To solve the above problem for large-scale data, a simple iterative procedure is adopted applying soft-thresholding and data consistency, simultaneously. This algorithm is popularly known by the name *projections over convex sets* (POCS) in the literature.

Table 1: Mutual coherence for different wavelet transforms

| Sl. No. | Wavelet transforms | μ |
|---|---|---|
| 1 | DWT | 0.9038 |
| 2 | DT-CoWT | 0.6611 |
| 3 | DD-DWT | 0.5693 |
| 4 | DD-DT-DWT | 0.7854 |

The basic steps of the proposed algorithm for our applications would be as follows:

---

**Algorithm 1: Proposed Algorithm**

1. **Model selection**
   $Y = R\phi\psi^T x = Ax$
   *If $Y$ frequency domain data, then $\phi = F$ and $R\phi = F_u$ ; $F_u$ =random partial Fourier matrix*
   *If $Y$ physical domain data, then $\phi = I$ and $R\phi = I_u$ ; $I_u$ =Id matrix with skipped diagonal elements*
2. **Zero-filling and Initialization**
   *Insert zeros for missing data in $Y$ and initialize: $\hat{X}_0 = Y$*
3. **Repeat**
   *If $Y$ is in frequency domain: apply IFFT to estimate: $\hat{x}_i = F^H \hat{X}_i$ otherwise: $\hat{x}_i = \hat{X}_i$*
4. **Apply soft-thresholding**
   $\hat{x}_i = S(\hat{x}_i, \lambda)$ *If $Y$ is in frequency domain: compute $\hat{X}_i = F \hat{x}_i$ otherwise go to the next step*
5. **Data consistency**
   $$\hat{X}_{i+1}[j] = \begin{cases} \hat{X}_i[j] & \text{if } Y[j] = 0 \\ Y[j] & \text{otherwise} \end{cases}$$
6. **Until convergence**
   $\|\hat{x}_{i+1} - \hat{x}_i\| < \varepsilon$

---

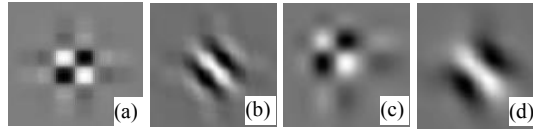

(a)      (b)      (c)      (d)

Figure 4: HH sub-band images for (a) DWT, (b) DT-CoWT, (c) DD-DWT, (d) DD-DT-DWT

## 4 Experimental Results

Simulations are carried out in the MATLAB environment run on a PC equipped with Intel i7 processor, 4GB of RAM. Overcomplete wavelet dictionaries are simulated using source codes available at [7, 8]. Results are obtained using test images collected from standard image database[1] and MRI images collected from[2]. All the results are averaged over many runs to avoid any bias.

### 4.1 Evaluation of mutual coherence (µ) between Fourier and overcomplete wavelet transforms

By applying the procedure in subsection 3.1, µ for different overcomplete wavelet transforms with the Fourier sensing matrix are calculated. Results are given in Table 1. DD-CoWT gives the least coherence approximations and they are most incoherent than other two overcomplete wavelet transforms in the frequency domain.

### 4.2 Image reconstruction in frequency domain

CS reconstruction is performed using Algorithm 1 using different wavelet transforms for a test image shown in Fig. 5 (a). Undersampling in the frequency domain is carried out using the mask in Fig. 5 (b). Least RMS error is observed for the DT-CoWT as shown in Table 2 and Fig. 6 (a). Although theoretical estimation gives the least value of mutual coherence for DD-DWT, yet the performance of DD-CoWT is better than DD-DWT for sparse signal reconstruction in the frequency domain. This may be because DD-CoWT is complex, nearly analytic, and highly directional compared to the DD-DWT. Reconstructed images in Figs. 7 (a)-(j) show that the DD-CoWT produces relatively less visible errors as observed in Fig. 7(b). All the algorithms converge after 20-25 iterations as shown in Fig. 6 (b).

### 4.3 Image reconstruction in physical domain

POCS algorithm is also used for image reconstruction in the physical domain. The mask in physical domain $R$ is directly applied to the test image in Fig. 8 (a). Results are observed after reconstruction of 20% discarded pixels and presented in Table 3 and Figs. 9 (a)-(j). Here again, DT-CoWT performs better than other transforms.

------------------------

1. The USC-SIPI Image Database, http://sipi.usc.edu/database/
2. The MRI data, https://people.eecs.berkeley.edu/~mlustig/CS.html

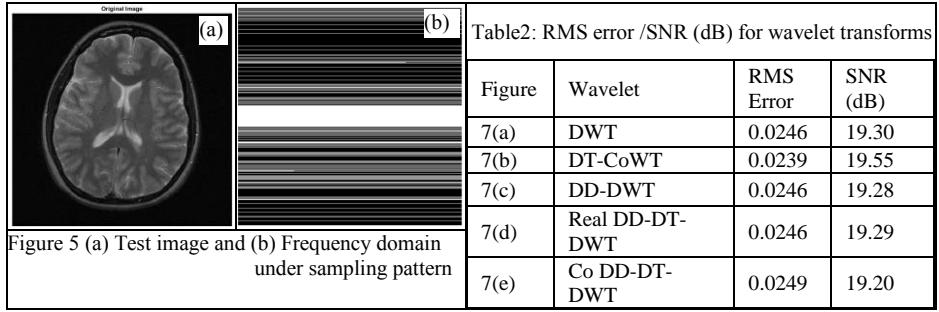

Figure 5 (a) Test image and (b) Frequency domain under sampling pattern

Table2: RMS error /SNR (dB) for wavelet transforms

| Figure | Wavelet | RMS Error | SNR (dB) |
|---|---|---|---|
| 7(a) | DWT | 0.0246 | 19.30 |
| 7(b) | DT-CoWT | 0.0239 | 19.55 |
| 7(c) | DD-DWT | 0.0246 | 19.28 |
| 7(d) | Real DD-DT-DWT | 0.0246 | 19.29 |
| 7(e) | Co DD-DT-DWT | 0.0249 | 19.20 |

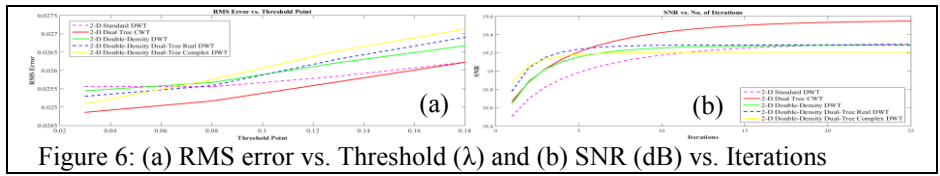

Figure 6: (a) RMS error vs. Threshold (λ) and (b) SNR (dB) vs. Iterations

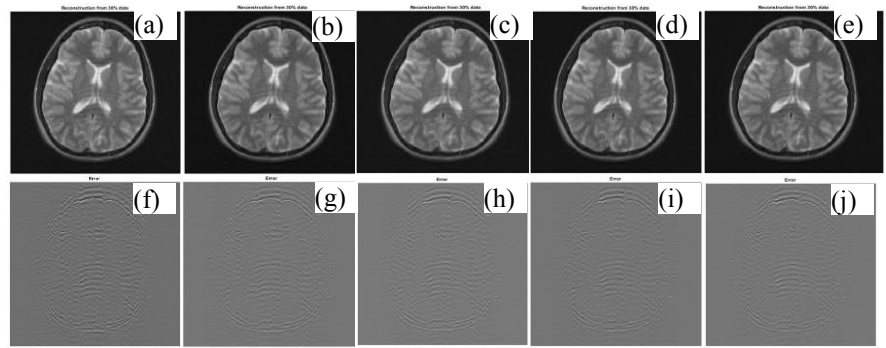

Figure 7: Reconstructed images using different wavelet transforms

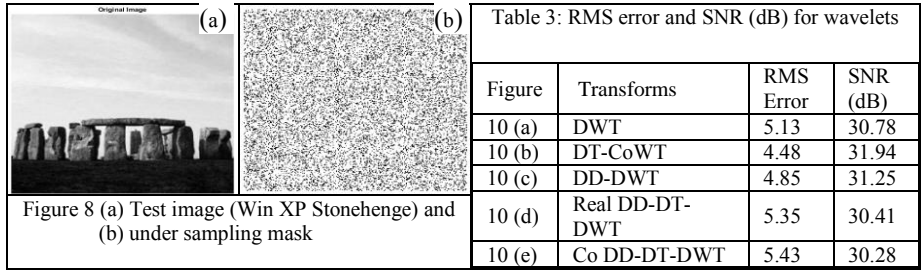

Figure 8 (a) Test image (Win XP Stonehenge) and (b) under sampling mask

Table 3: RMS error and SNR (dB) for wavelets

| Figure | Transforms | RMS Error | SNR (dB) |
|---|---|---|---|
| 10 (a) | DWT | 5.13 | 30.78 |
| 10 (b) | DT-CoWT | 4.48 | 31.94 |
| 10 (c) | DD-DWT | 4.85 | 31.25 |
| 10 (d) | Real DD-DT-DWT | 5.35 | 30.41 |
| 10 (e) | Co DD-DT-DWT | 5.43 | 30.28 |

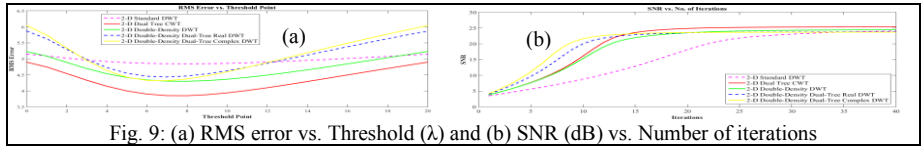

Fig. 9: (a) RMS error vs. Threshold (λ) and (b) SNR (dB) vs. Number of iterations

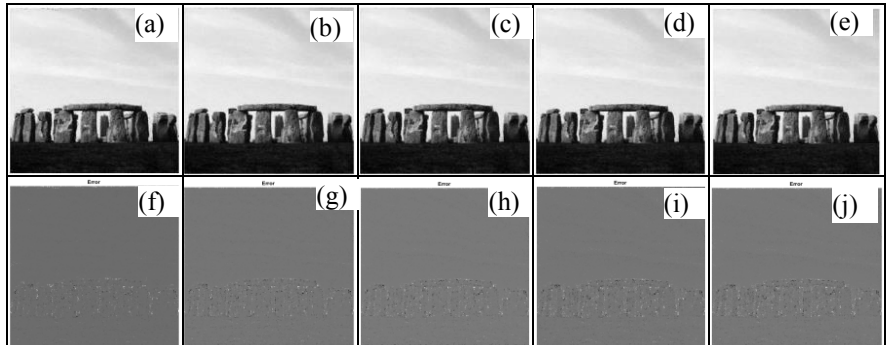
Fig. 10: Reconstructed images using different wavelet transforms

## 5 Conclusion

In this paper, experiments are carried out to find out the best overcomplete wavelet transform for CS reconstruction of images in frequency as well as spatial domains. Dual-tree complex wavelet transform performs better than other overcomplete wavelets in terms of compressive signal reconstruction. Results are evaluated on MRI and photographic images in terms of quantitative and visual analysis.